\documentclass[iop,apj]{emulateapj}
\usepackage{amsmath, amsthm, amssymb,natbib,multirow,rotating,ulem} 
\usepackage{natbib}
\usepackage{xspace}
\usepackage{xcolor}
\bibpunct{(}{)}{;}{a}{}{,} 
\usepackage[
  pdfstartview=FitH,
  linkcolor=blue,
  anchorcolor=blue,
  citecolor=blue,
  urlcolor=blue,
  colorlinks=true,
  breaklinks]{hyperref}
\newcommand{\figscalefact}{1.13}
\makeatletter
\newif\ifhbonecolumn
\@ifclasswith{emulateapj}{onecolumn}{\hbonecolumntrue}{\hbonecolumnfalse}
\newif\ifbaastex
\@ifclasswith{aastex}{iop}{\baastextrue}{\baastexfalse}
\makeatother

\newcommand{\figscale}[1]{
    \ifbaastex
        \epsscale{#1}
    \else
        \epsscale{\figscalefact}
    \fi
}

\shorttitle{Peculiar Outgassing of 17P/Holmes}
\shortauthors{Qi et al.}

\begin{document}

\title{Peculiar Near-Nucleus Outgassing of Comet 17P/Holmes During Its
  2007 Outburst}

\author{Chunhua~Qi$^1$, Michiel R. Hogerheijde$^2$, David Jewitt$^3$, 
Mark A. Gurwell$^1$, David J. Wilner$^1$    }

\affil{$^1$Harvard--Smithsonian Center for Astrophysics,
60 Garden Street, MS 42, Cambridge, MA 02138, USA}

\affil{$^2$Leiden Observatory, Leiden University, PO Box 9513, 2300 RA
  Leiden, The Netherlands}

\affil{$^3$Department of Earth, Planetary and Space Sciences and Department of Physics and Astronomy, University
  of California at Los Angeles, 595 Charles Young Drive East, Los
  Angeles, CA 90095, USA}

\begin{abstract}

We present high angular resolution Submillimeter Array observations of
the outbursting Jupiter family comet 17P/Holmes on 2007 October 26--29,
achieving a spatial resolution of
2.5$''$, or~$\sim$3000 km at the comet distance.  The observations resulted in
detections of the rotational lines CO 3--2, HCN 4--3, H$^{13}$CN 4--3,
CS 7--6, H$_2$CO 3$_{1,2}$-2$_{1,1}$, H$_2$S 2$_{2,0}$--2$_{1,1}$, and
multiple CH$_3$OH lines, along with the associated dust continuum at
221 and 349 GHz.  The continuum has a spectral index of 2.7$\pm$0.3,
slightly steeper than 
blackbody emission from large dust particles. From the imaging data,
we identify two components in 
the molecular emission. One component is characterized by a
relatively broad line width ($\sim$1 km s$^{-1}$ FWHM) exhibiting a
symmetric outgassing pattern with respect to the nucleus position. The
second component has a narrower line width ($<$0.5 km s$^{-1}$ FWHM)
with the line center red-shifted by 0.1--0.2 km s$^{-1}$ (cometocentric frame),
and shows a velocity shift across the nucleus
position with the position angle gradually changing
from $66\degr$ to $30\degr$ within the four days of observations.
We determine distinctly different CO/HCN ratios for each of the components.
For the broad-line component we find CO/HCN $<$ 7, while in the
narrow-line component, CO/HCN = 40 $\pm$ 5. We hypothesize that the
narrow-line component originates from the ice grain halo found in
near-nucleus photometry, believed to be created by sublimating
 recently released ice grains around the nucleus
during the outburst. In this
interpretation, the high CO/HCN ratio of this component
reflects the more pristine volatile composition of nucleus material
released in the outburst.

\end{abstract}


\section{Introduction}

Comets are thought to contain the most primitive material left over
from the planetesimal building stage of the Sun's protoplanetary disk. Their
physical and chemical composition provides an important link between
nebular and interstellar processes \citep{Ehrenfreund00,Mumma11}. 
A fundamental goal of cometary research is to understand the origin and
nature of cometary nuclei, in order to explore how they link back to
the protoplanetary disk. The most abundant constituent of a cometary
nucleus, water, is difficult to detect in ground-based observations
due to interference from the Earth's atmospheric water vapor.
However, many other molecules in cometary atmospheres can be detected,
and these can be used to trace the distribution and kinematics of gas
production from a comet. Of particular interest are parent molecules
(those sequestered in the cometary nucleus), which sublime directly
from the nucleus. Measurements of parent molecule production rates
can reveal relative abundances in the nucleus itself and therefore
provide information about the conditions under which the comet formed.

There are at least two distinct long-term reservoirs for
cometary nuclei, the Oort Cloud and the Kuiper Belt.  
Measuring the composition from each of the reservoirs may shed light
on their origins in the protoplanetary disk from which they formed.
Radio interferometers possessing broad bandwidth coverage 
coupled with high spectral and spatial resolution can be used to
examine spatial differences among parent molecules to probe the
heterogeneity in the coma and nucleus (e.g., Blake et
al. 1999). Simultaneous sampling of multiple parent molecules can test
variations of abundances and abundance ratios associated with changes
in the active area (Mumma \& Charnley 2011).

However, it is not straightforward to determine the composition of a
comet without directly sampling the nucleus. Instead, we typically
measure the coma emission of parent and daughter species released
from the nucleus. This technique, while powerful, presents at least
two problems. The first is that it is usually difficult to obtain spatial
information on
multiple species simultaneously, which is critical for the determination of abundance ratios due to the variable
nature of comets (e.g., outflow asymmetries and jet activity). The second
is that the chemical composition in the coma is not necessarily
representative of the composition of the nucleus. For example,
chemical reactions in the coma can affect the coma 
composition (e.g., HNC is likely to be formed in situ via chemistry in
the inner coma (Mumma \& Charnley 2011)).  More seriously, thermal
processing of the comet surface is 
expected to lead to preferential volatile loss, resulting in a
gradient in composition with depth beneath the physical surface.
Sublimation then proceeds from a range of depths, depending on the
volatility of the ices, their permeability to gas flow, and the
progression of thermally conducted heat moving into the interior.  The
gases escaping from the heated surface may not be representative of
the bulk composition of the ices measured deep in the nucleus.  

Jupiter family comets likely originated in the Kuiper Belt reservoir,
although the precise source region within the Kuiper belt remains
unclear \citep{Volk08}. They have 
short orbital periods and average residence times in the inner solar
system of $\sim$0.4 Myr \citep{Levison94}. Volatile loss in previous
orbits makes it especially difficult to determine their initial
composition from observations of their comae; they can be very hard to 
observe because of their usually weak emission, and repeated
  passes near the Sun deplete volatiles from the surface and
  near-surface of the nucleus. However, a cometary outburst 
potentially provides the opportunity to release fresh, minimally altered
material from the interior of the nucleus.  The unexpected outburst of
Jupiter Family comet 17P/Holmes in 2007 provided just such
an opportunity to detect and study fresh material from the
  comet's deeper layers. 
  
  In this paper, we present Submillimeter Array
(SMA) observations of the outburst of 17P/Holmes, in which the
  brightness increased from  apparent magnitude $ V \sim$17 on UT 2007
  October 23.3 \citep{Hsieh10} to $V \sim$ 2.0 on UT 2007 October
  25.0 \citep{Li11}, corresponding to the sudden ejection of (2 to
  90)$\times$10$^{10}$ kg of material \citep[e.g.][]{Ishiguro10,Li11}.
  We use simultaneous 
imaging of HCN and CO from the outburst to identify two different
components in the molecular emission. These components 
exhibit drastic differences in volatile composition as
  measured by CO/HCN, suggesting that one component samples pristine
  material from deep in the nucleus.

\section{Observations}

\begin{figure}[t]
\figscale{0.5}
\plotone{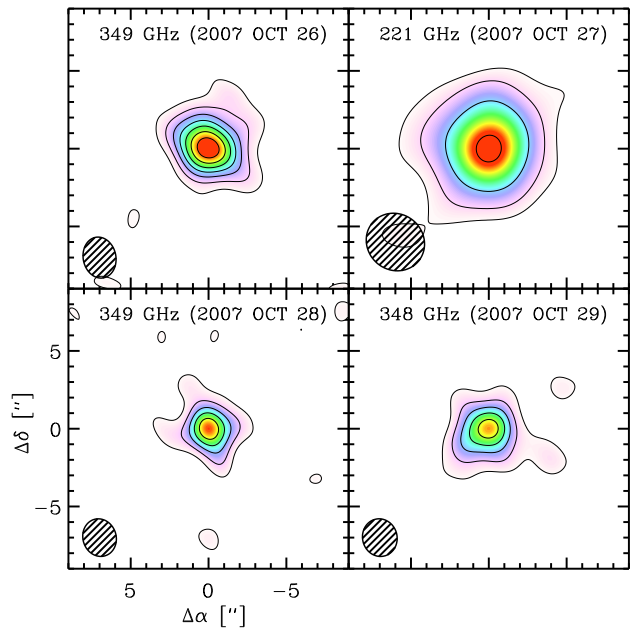}
\figcaption{Interferometric maps of the (sub)mm continuum emission of
comet 17P/Holmes obtained from 26 to 29 October 2007 UT with the
SMA. The contours are 2.5 mJy~Beam$^{-1}$ (1$\sigma$) $\times
  [2,4,6,8,10,12,14]$ (for October 26); 0.55 mJy~Beam$^{-1}$ (1$\sigma$) $\times
  [3,6,12,24]$ (for October 27); 2.0 (1$\sigma$) $\times
  [2,4,6,8,10]$ (for October 28); 1.8 (1$\sigma$) $\times
  [2,4,6,8,10]$ (for October 29). \label{fig:cont}
}
\end{figure}

Observations of 17P/Holmes were made on
  four consecutive days between 2007 October 26th and 29th using the
SMA eight antenna interferometer located atop Mauna Kea, Hawaii. 
  The SMA operates in the 1.3 and 0.86 mm atmospheric
  transmission windows which cover rotational transitions from a
  variety of important cometary volatile species, such as HCN, CO and
  CS.  The SMA was used in the compact configuration, resulting in
a spatial resolution of 2.5$''$ at 349 GHz and 4$''$ at 221 GHz,
corresponding to 2950 and 4730 km in linear scale at the Earth-comet
distance of 1.63 AU. The primary beams are 31.2$''$ and 49.8$''$ FWHM,
respectively. The SMA operated in a double-sideband (DSB) mode
with an intermediate frequency band of 4--6 GHz which was sent over
fiber optic transmission lines to 24 overlapping digital correlator
bands (or ``chunks'') covering the 2 GHz spectral
window. Phase-switching techniques were used to separate and
simultaneously recover both sidebands, resulting in 4 GHz of total
coverage (2 GHz per sideband). Doppler-tracking of 17P/Holmes was 
performed so that the rapidly changing velocity of the comet,
as defined by available ephemerides at the time of
observations (specifically, JPL orbital solution \#K077/6, HORIZONS
System), was compensated for in real time, so the velocity reported
here is under the cometocentric velocity frame. 
At all frequencies, simultaneous dust continuum measurements
were made using the line-free chunks. Tables 1 and 2 summarize the
observational parameters for the continuum and spectral line setups
used for each date of observation. In particular, observations on 
  October 26, 28, and 29 focused on HCN and either CO or CS near 349
  GHz, while those from October 27 aimed to detect H$_2$CO, H$_2$S,
  CH$_3$OH and the deuterated species HDO and DCN near 221 GHz. 

Calibration of visibility phases and amplitudes was achieved with 
periodic snapshots of the quasar 0359+509 at 20 minute intervals. 
Measurements of Uranus provided an absolute scale for calibration of
the flux densities. The derived mean flux densities of 0359+509 at the
time of the observations were 2.05 Jy at 349 GHz and 3.25 Jy at 221
GHz. All data were phase- and amplitude-calibrated using
the MIR software
package \footnote{http://www.cfa.harvard.edu/$\sim$cqi/mircook.html}.
Continuum and spectral line maps were then generated and CLEANed using
the MIRIAD software package.

\section{Results}

\begin{figure*}[htbp]
\includegraphics[width=3in]{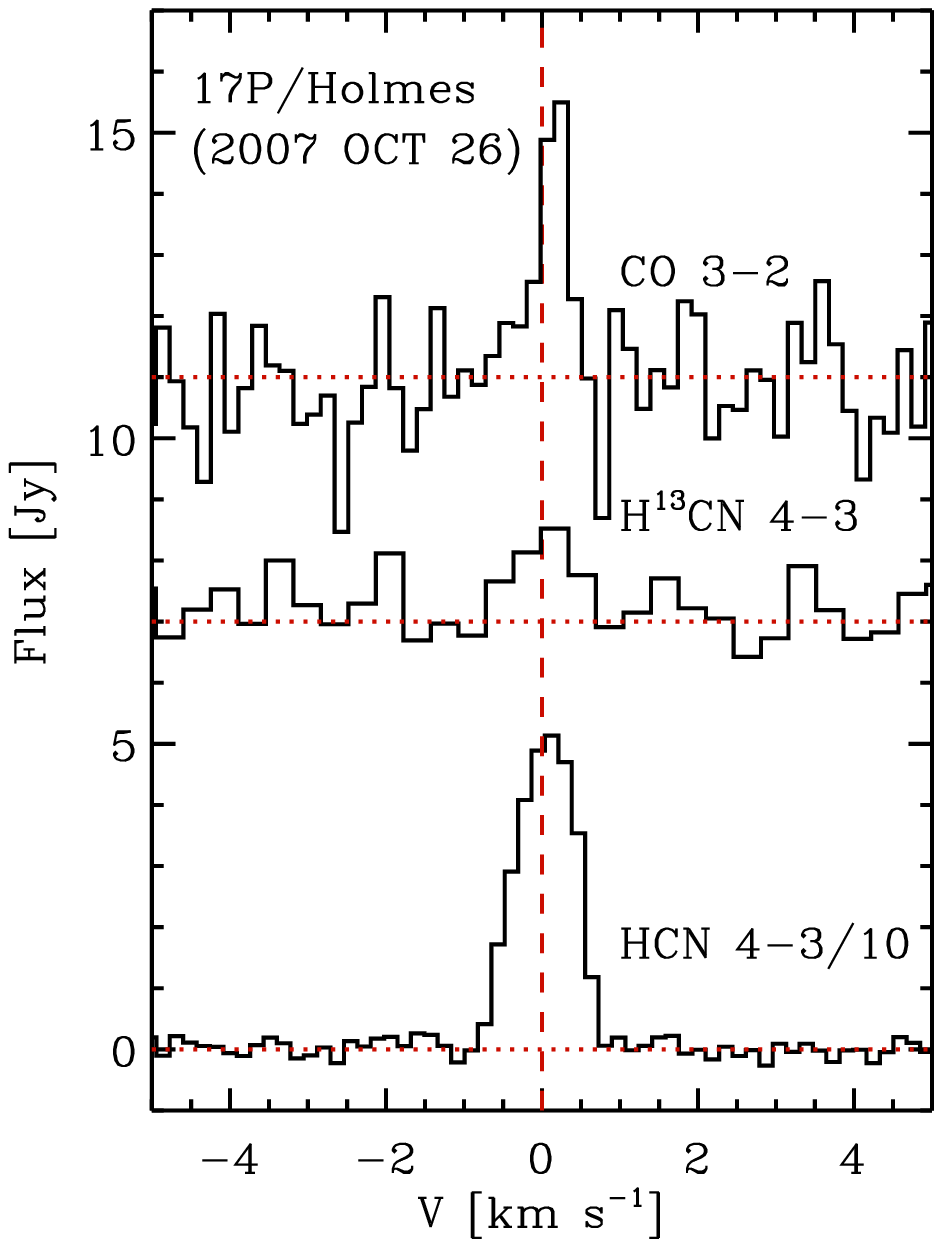}
\includegraphics[width=3in]{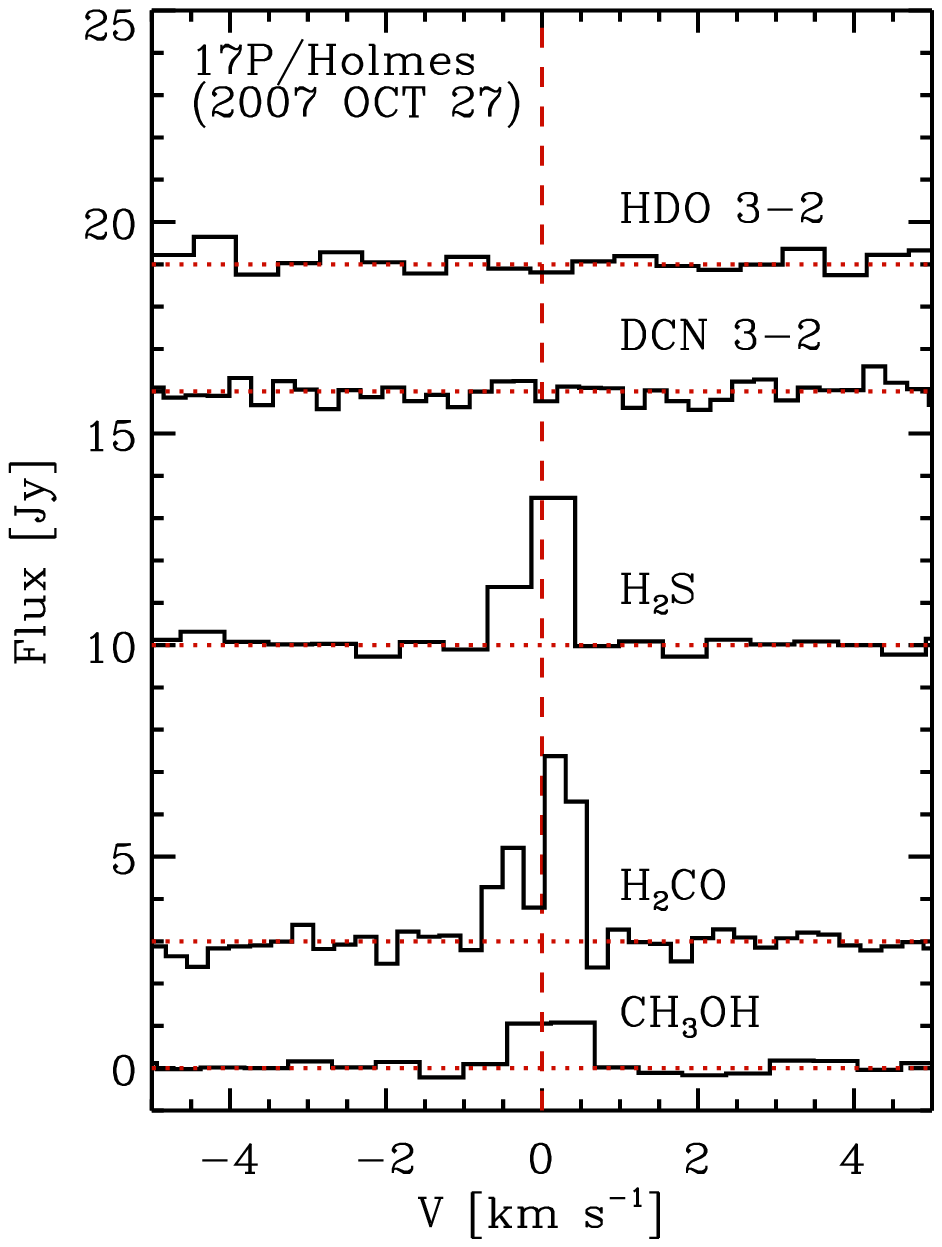} \\
\includegraphics[width=3in]{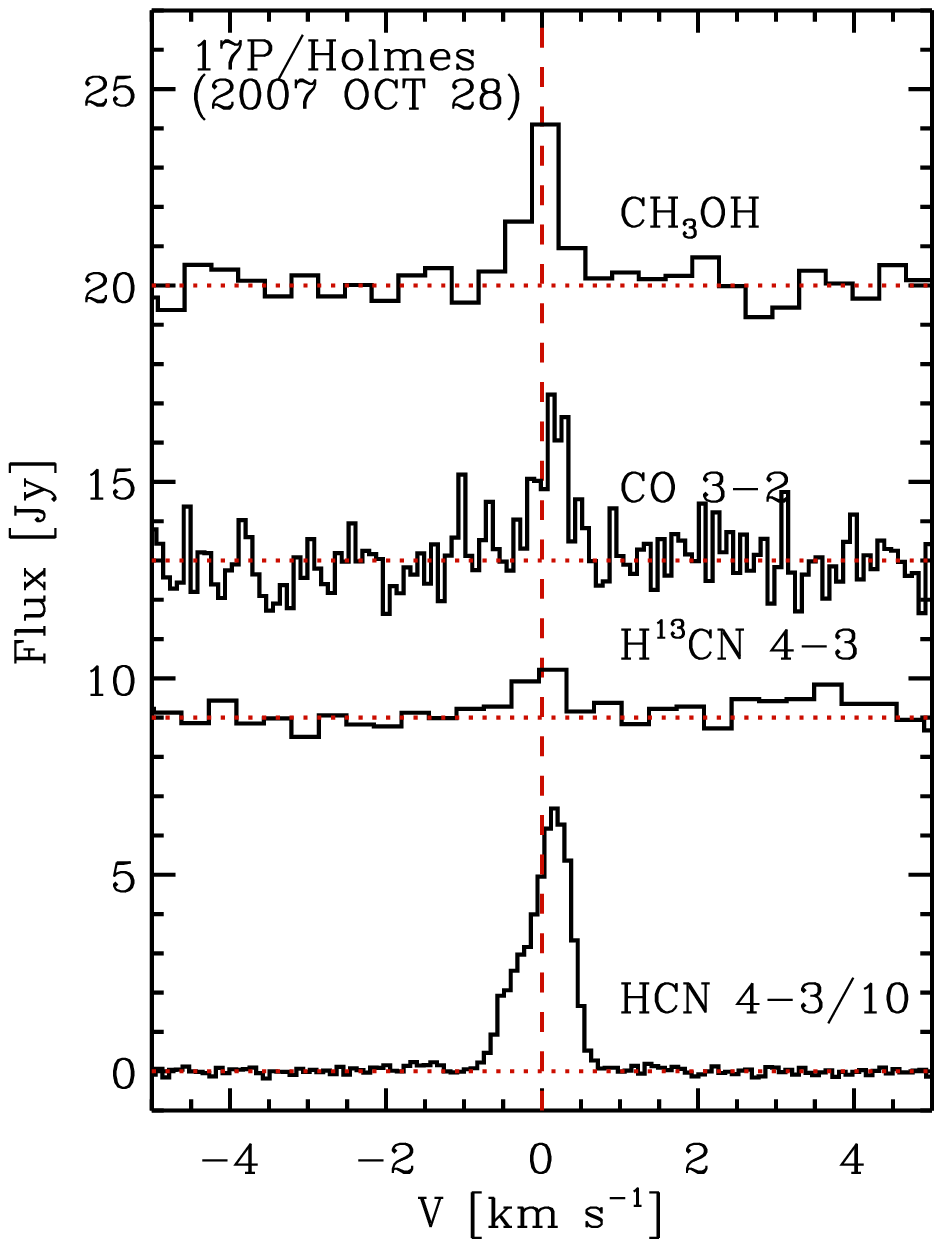}
\includegraphics[width=3in]{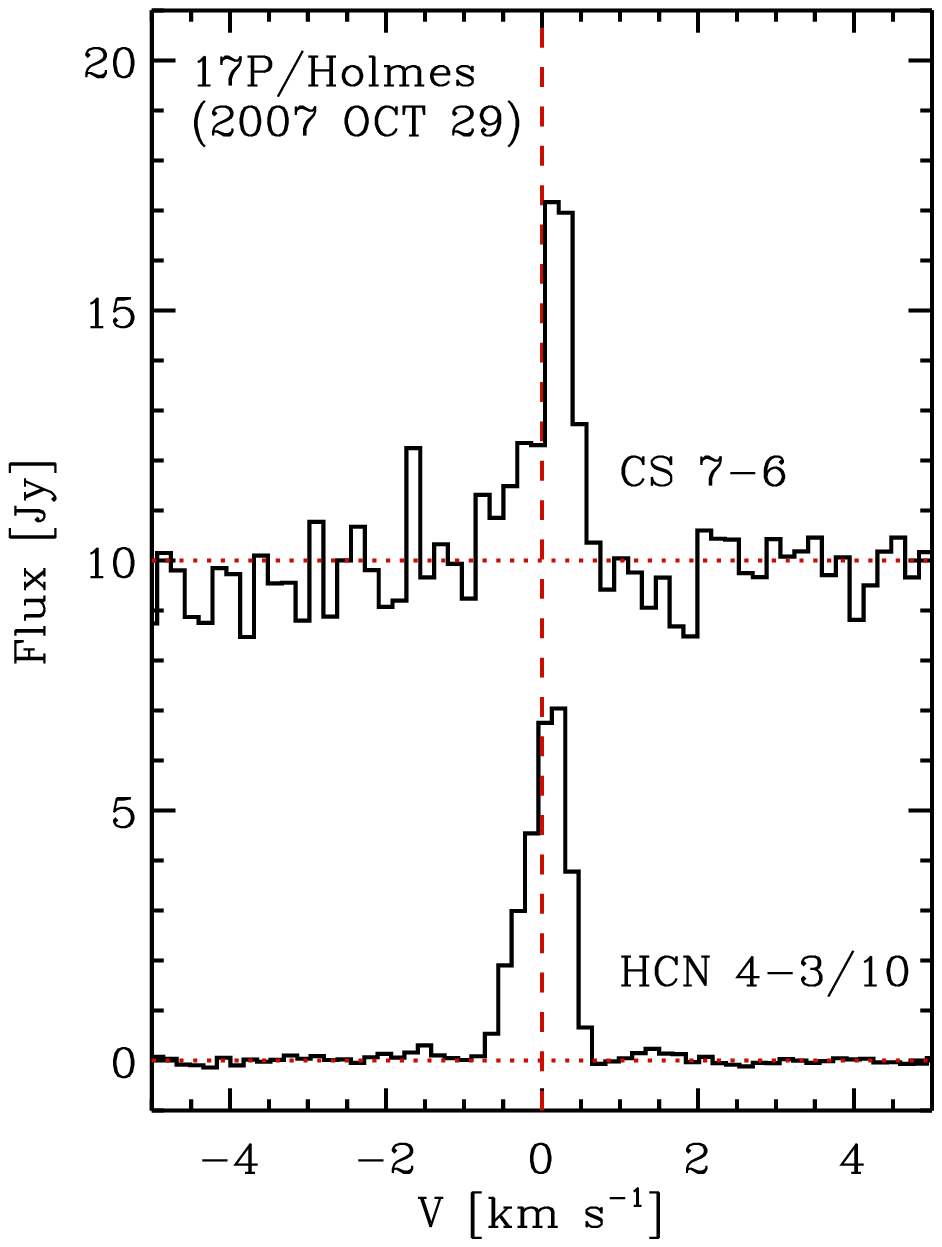}
\caption{Spatially integrated spectra at the peak continuum position
  of 17P/Holmes. The fluxes are averaged over the emission areas.
  All spectra are plotted in a cometocentric velocity frame. 
  The vertical dashed line mark the cometocentric velocity of 0 km~s$^{-1}$.
\label{fig:spec} }
\end{figure*}

Continuum emission at 221 and 349 GHz is readily detected by the
SMA. Figure~\ref{fig:cont} shows (sub)mm continuum emission maps of
comet 17P/Holmes obtained from 26 to 29 October 2007 and integrated
flux densities are reported in Table~\ref{tab:cont}.  The (sub)mm
continuum emission probes the thermal radiation of mm-sized dust
particles in the coma and the peak of the emission is taken to reveal the
location of the nucleus \citep[e.g.][]{Jewitt92, Blake99, Qi01}.

\begin{deluxetable*}{lcccccc}
\tablewidth{0pt}
\tablecaption{Observational parameters for continuum observations \label{tab:cont}}
\tablehead{
\colhead{Date(UT)} &
\colhead{$\lambda$(mm)}&\colhead{Frequency(GHz)}&\colhead{Beam}&\colhead{PA}&\colhead{Flux
  (mJy)} &\colhead{Offsets ($''$)} }
\startdata
2007 October 26.3-26.7 & 0.86 & 349 & $2\farcs7 \times 2\farcs1$ & 11.2$^\circ$ &
96.4$\pm$2.3 & (0.92,-0.26) \\
2007 October 27.3-27.7 & 1.35 & 221 & $3\farcs9 \times 3\farcs6$ & 40.8$^\circ$ &
30.3$\pm$0.6 & (0.59,-0.20) \\
2007 October 28.3-28.7 & 0.86 & 349 & $2\farcs4 \times 2\farcs1$ & 16.9$^\circ$ &
40.9$\pm$1.8 & (0.85,-0.26) \\
2007 October 29.3-29.7 & 0.86 & 348 & $2\farcs4 \times 2\farcs2$ & 13.7$^\circ$ &
48.8$\pm$1.4 & (0.83,-0.17) \\
\hline
\enddata
\end{deluxetable*}

Table~\ref{tab:line} lists the observational parameters and
  resulting detections for the spectral line observations,
  while in Figure~\ref{fig:spec} the spectra from
  several molecular species are displayed for each day. These
  spectra are all extracted from a 10$''$ square box centered on the
nucleus, corresponding to the maximum area with significant
emission. We note that the decrease of the continuum flux from October 26 to
29 is not smooth as the continuum flux on the 29th is slightly higher than
that on the 28th. This is also accompanied by a small but significant
increase in the velocity-integrated intensity level of HCN 4--3 as
reported in Table 2 from the 28th to 29th. This is probably caused by the
intermittent release of both big dust grains and volatiles sublimating
from the small icy grains near the nucleus by the outburst.

\begin{deluxetable*}{lcccccc}
\tabletypesize{\footnotesize}
\tablecolumns{7} 
\tablewidth{0pt} 
\tablecaption{Observational parameters for line observations \label{tab:line}}
\tablehead{\colhead{Lines}              &
           \colhead{Rest Frequency}     &
	   \colhead{Beam}               &
	   \colhead{PA}                 &
	   \colhead{Channel spacing}   &
           \colhead{FWHM\,\tablenotemark{a}} &
	   \colhead{Int. Intensity.\,\tablenotemark{b}}    \\
	   \colhead{}                   &
	   \colhead{(GHz)}              &
	   \colhead{($''$)}             &
	   \colhead{($^\circ$)}         &
	   \colhead{(km\,s$^{-1}$)}      &
	   \colhead{(km\,s$^{-1}$)}      &
	   \colhead{(Jy\,km\,s$^{-1}$)} }
\startdata 
\cutinhead{UT: 2007 October 26.3-26.7}	   
HCN 4--3 & 354.505 & 2.7$\times$2.1 & 12.9 & 0.17 &0.89
& 52.09[0.22] \\
H$^{13}$CN 4--3 & 345.340 & 2.6$\times$2.0 & 9.6& 0.35 & 0.92
& 1.32[0.22] \\
CO 3--2 & 345.796 & 2.6$\times$2.0 & 9.6 & 0.18 & 0.42
& 2.27[0.21] \\
\cutinhead{UT: 2007 October 27.3-27.7}
H$_2$CO 3$_{1,2}$--2$_{1,1}$ & 225.698 & 3.8$\times$3.2 & 49.6& 0.27&0.72
& 3.03[0.14] \\
HDO 3$_{1,2}$--2$_{2,1}$ & 225.897 & 3.8$\times$3.2 & 49.6& 0.54 & --
& [0.14] \\ 
DCN 3--2 & 217.239 & 4.2$\times$3.9 & -39.1 & 0.28 & --
& [0.14] \\
H$_2$S 2$_{2,0}$--2$_{1,1}$ & 216.710 & 4.2$\times$3.9 & -39.1 & 0.56 & --
& 1.36[0.14] \\
CH$_3$OH 5$_{1,0}$--4$_{2,0}$ & 216.946 & 4.2$\times$3.9 & -39.1& 0.56&--
& 1.34[0.14] \\
\cutinhead{UT: 2007 October 28.3-28.7}
HCN 4--3 & 354.505 & 2.5$\times$2.1 & 19.4& 0.086 & 0.69 
& 48.79[0.16] \\
H$^{13}$CN 4--3 & 345.340 & 2.4$\times$2.1 & 14.2 & 0.35 & 0.68 
& 1.07[0.16]\\
CO 3--2 & 345.796 & 2.4$\times$2.1 & 14.2 & 0.088 & 0.50
& 1.75[0.15] \\
CH$_3$OH 13$_{0,0}$--12$_{1,0}$ & 355.603 & 2.5$\times$2.1 & 19.4& 0.34&--
& 2.40[0.16] \\
\cutinhead{UT: 2007 October 29.3-29.7}
HCN 4--3 & 354.505 & 2.4$\times$2.1 & 15.6 & 0.17 &0.65
& 50.16[0.21] \\
CS 7--6 & 342.883 & 2.4$\times$2.2 & 11.1 & 0.18 &0.46
& 4.45[0.21]\\
\enddata
\tablenotetext{a}{ 
 Fitted with a Gaussian. Might not reflect the real line width.} 
\tablenotetext{b}{Integrated over 10$''$ box. }
\end{deluxetable*}

For each day, we were able to cover several species within the 4 GHz
bandwidth (both upper and lower sidebands). This 
 provides the opportunity to compare the extent of the emission from
different species observed simultaneously in the coma during the
outburst. On October 27, the SMA was tuned near 221 GHz (1.4mm)
  in order to detect H$_2$CO, H$_2$S and CH$_3$OH, and search for the deuterated
  species HDO and DCN (see Table~\ref{tab:line}). Observations near 349 GHz
  (0.86mm) of HCN with either CO or CS were obtained on October 26, 28
  and 29.  While several species at 1.4mm were detected and the
  integrated intensities are reported in Table~\ref{tab:line}, the velocity
  resolution was insufficient for a detailed kinematic study, so in
  this paper we focus on the HCN and CO/CS imaging and defer the
  analysis of
  H$_2$CO, CH$_3$OH and H$_2$S to future work.

\begin{figure*}[htbp]
\figscale{0.6}
\plotone{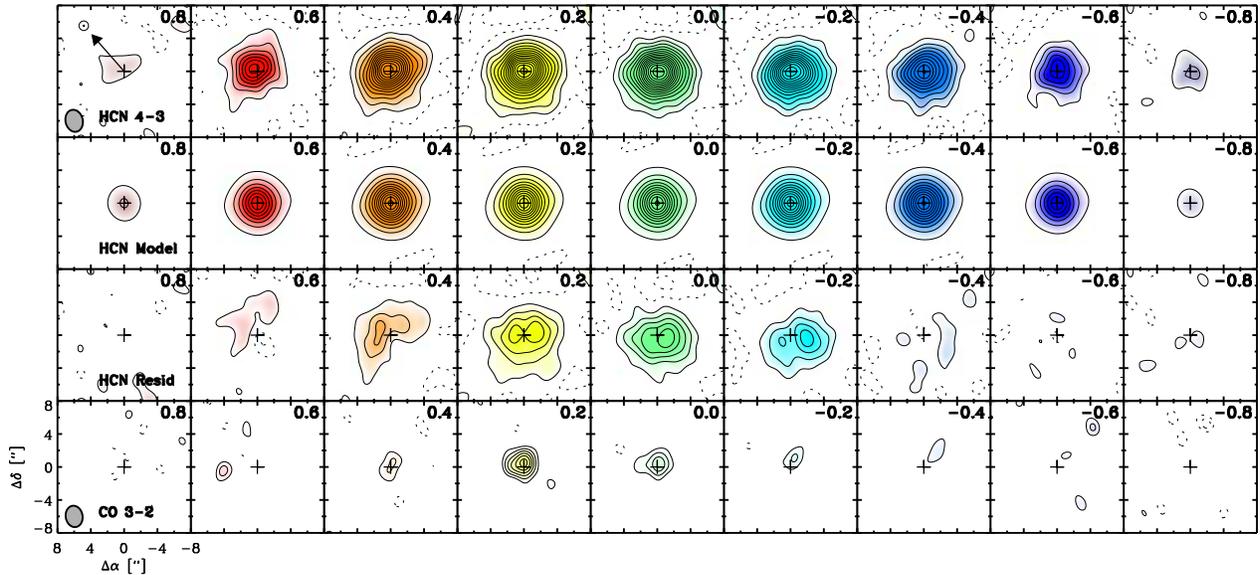}
\figcaption{ SMA observations of molecular emission from Comet 17P/Holmes
  on October 26th, 2007, 2 days after the onset of outburst. The
  residual of the HCN 4--3 
  emission (third row), obtained by subtracting the best-fit
  symmetric outgassing model (second row) from the data (first
  row), is likely to created by sublimating recently released ice
  grains around the nucleus during the outburst. 
  The extent of the CO 3--2
  emission (fourth row) is consistent with the residual of HCN
  emission, indicating a higher volatile content in this
  component. For HCN emission, the contours start with 2$\sigma$ and
  each following contour step is 3$\sigma$ (1$\sigma$=0.3
  Jy~Beam$^{-1}$). For CO emission, the contours start with 2$\sigma$
  and step in 1$\sigma$. The cross marks the position of the nucleus
  and arrows show the direction of the Sun. North is the positive
  Y-axis and East is the positive X axis. \label{fig:oct26}
}
\end{figure*}

To better compare the emission of the different species, we
regenerated channel images of the emission lines on a common velocity
grid. Figures~\ref{fig:oct26}, \ref{fig:oct28}, and \ref{fig:oct29}
present the resampled HCN 4--3, CO 3--2, and CS 7--6 emission.
The line width of HCN
4--3 is much broader than those of CO 3--2 or CS
7--6 for each day, and the emission of HCN is also more extended
spatially than that of CO and CS. Most of the HCN emission
appears symmetric, consistent with expectations for isotropic
outgassing. However, there are some deviations from this symmetric
pattern. Some HCN emission (for
  example near +0.4 km s$^{-1}$)
  extends toward the northeast.  In contrast, the CO
3--2 and CS 7--6 emission is significantly narrower
in velocity width, and for both species the
emission shifts position with velocity, showing a
  gradient across the continuum (nucleus) position. The peak emission 
from both lines is around the nucleus position, but about 0.1--0.2
km~s$^{-1}$ red-shifted (cometocentric frame). 

\section{Analysis}

\subsection{Continuum}
Continuum measurements at 0.86 mm (Table \ref{tab:cont}) from UT October 26, 28 and 29 show fading by a factor $\sim$2, roughly corresponding to an exponential decay with e-folding time of $\sim$4 days.  By interpolation, the  flux density at wavelength $\lambda_{0.86}$ = 0.86 mm on UT October 27 is estimated to be $S_{0.86}$ = 68$\pm$7 mJy, which compares with the flux density at $\lambda_{1.35}$ = 1.35 mm on this date of $S_{1.35}$ = 30.3$\pm$0.6 mJy.  The different beam sizes at 0.86 mm and 1.35 mm sample different volumes of the near-nucleus dust coma, requiring a correction.  We assume that the dust density varies with the inverse square of the distance from the nucleus, in which case the quantity of dust intercepted by an aperture  varies in proportion to $\phi$, the angular radius of the aperture.  The effective aperture size is given by the projected beam which, for our data, is elliptical.  We take the effective beam radius as $\phi = (\phi_{max} \phi_{min})^{1/2}$, where $\phi_{min}$ and $\phi_{max}$ are the beam ellipse maximum and minimum radii from Table \ref{tab:cont}.  We find $\phi_{0.86}$ = 2.3\arcsec~and $\phi_{1.35}$ = 3.7\arcsec. Then, the spectral index of the continuum of 17P in the wavelength range from 0.86 to 1.35 mm is given by

\begin{equation}
\alpha = \frac{log\left[(S_{1.35}/S_{0.86})(\phi_{0.86}/\phi_{1.35})\right]}{log(\lambda_{1.35}/\lambda_{0.86})}
\end{equation}

\noindent  By substitution, we obtain $\alpha$ = 2.7$\pm$0.3, where
the quoted uncertainty mainly reflects the error in $S_{0.86}$ on UT
October 27 introduced by the interpolation from adjacent dates.
Evidently, $\alpha$ is slightly larger than  the expected
blackbody spectral index in the Rayleigh-Jeans regime, namely $\alpha$ = 2.  However, it is 
consistent with earlier measurements of dust in comet C/Hale-Bopp,
which gave $\alpha$ = 2.60$\pm$0.13 (Jewitt and Matthews 1999).
Values of $\alpha$ near 2 indicate that the continuum cross-section is dominated by optically large dust
particles ($2\pi a/\lambda > 1$, corresponding to particle radii $a \gtrsim$ 200 $\mu$m). 

We used Equation (6) of Jewitt and Matthews (1999) to estimate the mass loss rate from $S_{1.35}$  (Table \ref{tab:cont}), obtaining $dM/dt$ = 2.6$\times$10$^{6}$ kg s$^{-1}$ on UT October 27.  This value, about 1000 times larger than typical of Jupiter family comets at the same heliocentric distance, may be compared with the peak mass loss rate three days earlier on UT 2007 October 24.5, estimated from optical continuum data as being up to 1.4$\times$10$^7$ kg s$^{-1}$ (Li et al.~2011).  On the other hand, the (gas) production rate five days later, on UT 2007 November 01, had fallen to 2$\times$10$^4$ kg s$^{-1}$ (Schleicher 2009).  Even accounting for the decay of the brightness of 17P between October 27 and November 01, it appears that 17P was an extremely dust-rich comet.

\subsection{Symmetric outgassing model }

\begin{sidewaysfigure}
\includegraphics[width=9in]{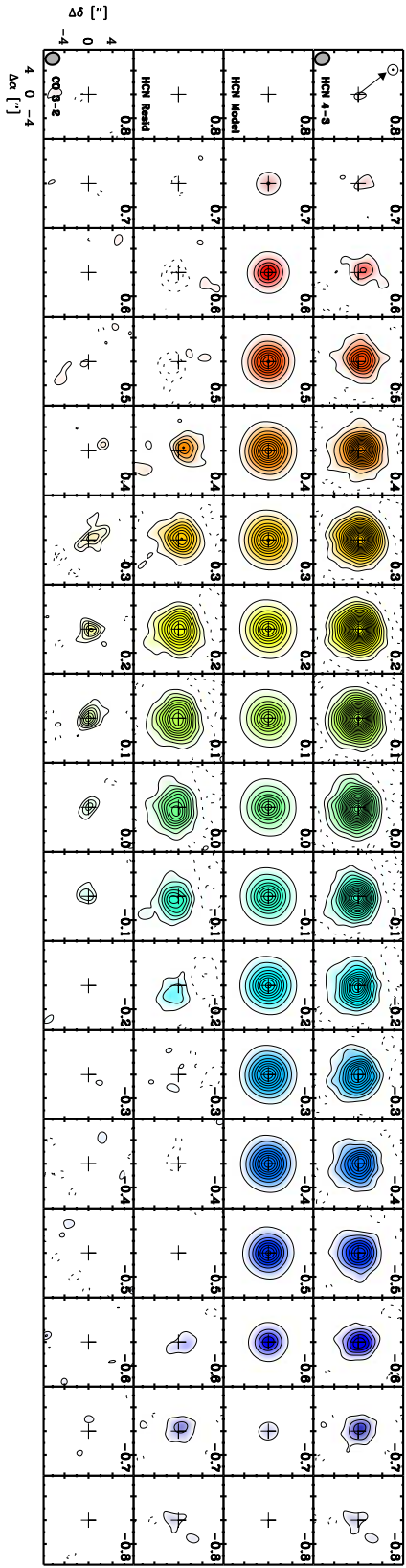}
\caption{ Same  as figure~\ref{fig:oct26}, but for October
  28th, 2007. \label{fig:oct28}
}
\end{sidewaysfigure}

\begin{figure*}[htbp]
\figscale{0.6}
\plotone{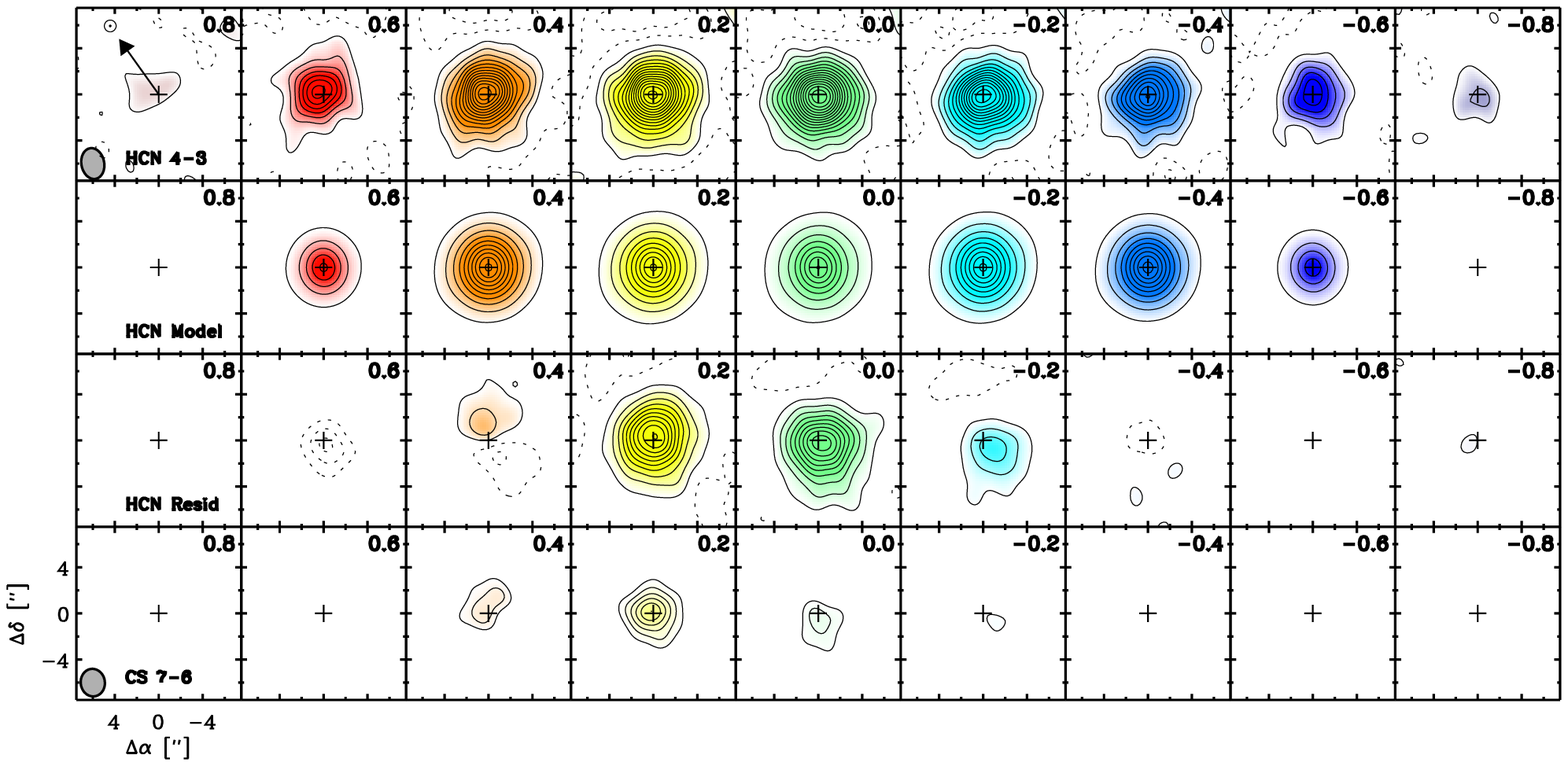}
\figcaption{ Same as figure~\ref{fig:oct26}, but for October 29th,
  2007. For CS emission, the contours start with  2$\sigma$
  and step in 1$\sigma$ (1$\sigma$=0.3 Jy~Beam$^{-1}$). \label{fig:oct29}  
}
\end{figure*}

Figures~\ref{fig:oct26}, \ref{fig:oct28}, and \ref{fig:oct29}
suggest that much of the HCN emission toward 17P/Holmes
originated from symmetric outgassing. This emission typically has a
relatively large linewidth (FWHM $\sim$ 1 km~s$^{-1}$). There is
also clear evidence that some HCN emission does not follow this
symmetric pattern. The CO 3--2 and CS 7--6 emission is dominated by
emission asymmetric with respect to the systemic velocity of the comet (0 km~s$^{-1}$ in the cometocentric frame), and 
shows smaller line widths of FWHM $<$ 0.5 km~s$^{-1}$. Is there a
relation between the asymmetries in the HCN line and the narrow CO and
CS emission? To answer this question requires
detailed modeling of the HCN emission. The modeling
analysis consists of the following steps. First, we construct a
symmetric outgassing model for HCN.  Second, we subtract the symmetric
outgassing model from the data, and compare the subtracted HCN
residual with the CO 3--2 and CS 7--6 emission to determine if they
share the same origin. Finally, we investigate any compositional
differences between the symmetric and asymmetric parts of the coma.

For a first-order analysis of the distribution of HCN in the coma, we
assume that the density in the coma drops with distance from the
nucleus, which is the only source for HCN.  The
distribution of density as a function of radius can
be approximated using the Haser model with a constant expansion velocity
v$_{\rm exp}$. The modeling procedure is similar to
  that presented in Hogerheijde et al.~(2009), adapted to the
  geometrical circumstances of 17P/Holmes.  We calculate line  
formation and molecular
excitation in the coma using an adapted version of the Accelerated
Monte Carlo code by Hogerheijde \& van der Tak (2000), in which we
consider excitation via collisions with water and electrons,
spontaneous and induced emission of line photons, and absorption of
line and continuum photons, including those of the solar infrared
spectrum (see Section 4.2 of Hogerheijde et al.~(2009) for details of
molecular excitation calculation).  For the main collisional partner,
we adopt a water production rate Q(H$_2$O) = 7, 4, and
3$\times$10$^{29}$ mol~s$^{-1}$, appropriate for October 26th, 28th, and
29th \citep{DelloRusso08}. For the gas temperature, we adopt a constant
kinetic temperature of 45 K, based on the results of Bockel\'ee-Morvan
et al.~(2008).

\begin{deluxetable}{cccc}
\tablewidth{0pt}
\tablecaption{Spectroscopic values for HCN 4--3 hyperfine components \label{tab:hcnhfs}}
\tablehead{
\colhead{J F} & \colhead{J' F'}&\colhead{ Frequency (GHz)}&\colhead{
Rel. Intensity}  }
\startdata
4 4 & 3 4 & 354.5038689 & 0.021 \\
4 3 & 3 2 & 354.5053670 & 0.24 \\
4 4 & 3 3 & 354.5054778\,\tablenotemark{a} & 0.31 \\
4 5 & 3 4 & 354.5055234 & 0.41 \\
4 3 & 3 4 & 354.5058468 & 5$\times$10$^{-4}$ \\
4 3 & 3 3 & 354.5074558 & 0.021 \\
\hline
\enddata
\tablenotetext{a}{Reference frequency}
\end{deluxetable}

HCN has a hyperfine structure due to the nuclear quadruple moment of
$^{14}$N. We adopt the six HCN hyperfine components of the 4--3 transition and
their relative intensities from the JPL Molecular Spectroscopy database 
 listed in the Table~\ref{tab:hcnhfs}.
Relative populations between the hyperfine levels are
assumed to be in LTE. The HCN emission is not very optically
thick. For example on October 28th, the integrated intensity of the
central four hyperfine components is 46.30 Jy~km~s$^{-1}$, and the
outer hyperfine component 4$_3$--3$_3$ (well separated from the
others) is 1.25 Jy~km~s$^{-1}$. So the 
measured ratio of this component over the central four combined is
0.026$\pm$0.003, close to the expected ratio of 0.022 for optically
thin emission. After construction of the emission model, we multiply
it by the appropriate primary beams
and resample with the $uv$ data using the actual SMA antenna positions
to generate model visibilities. 
To find the best-fitting description of the symmetric emission,
we only consider emission outside the cometocentric velocity range of
$-0.2$ to $+0.4$ km s$^{-1}$, corresponding to the range outside
where the narrow CO and CS emission dominates.  Using the comet model
and the excitation calculation described above, we compute a grid of
synthetic HCN visibility datasets over a range of Q$_{\rm HCN}$ and
V$_{\rm exp}$ values and compare with the observations. The best-fit
model is obtained by minimizing the $\chi^2$, the weighted difference
between the data and the model with the real and imaginary part of the
complex visibility measured in the $(u,v)$-plane sampled by the HCN
observations. Our analysis is performed in the visibility domain
rather than the image domain, to avoid the non-linear effects of the
imaging and deconvolution process.

\begin{deluxetable}{lcc}
\tablewidth{0pt}
\tablecaption{HCN symmetric outgassing model fitting result\,\tablenotemark{a} \label{tab:fitting}}
\tablehead{
\colhead{Date} &
\colhead{v$_{\rm exp}$(km s$^{-1}$)}&\colhead{Q$_{\rm HCN}$(10$^{26}$ mol s$^{-1}$)} }
\startdata
2007 October 26 & 0.46[0.01] & 12.5[1.0] \\
2007 October 28 & 0.44[0.01] & 8.0[0.3] \\
2007 October 29 & 0.38[0.01] & 6.3[0.2] \\
\hline
\enddata
\tablenotetext{a} { Symmetric outgassing model fitting on velocity range outside of $-$0.2
  to $+$0.4 km $^{-1}$, i.e. not including the asymmetric narrow-line
  component of the HCN emission.}
\end{deluxetable}

\begin{figure}
\includegraphics[width=3in]{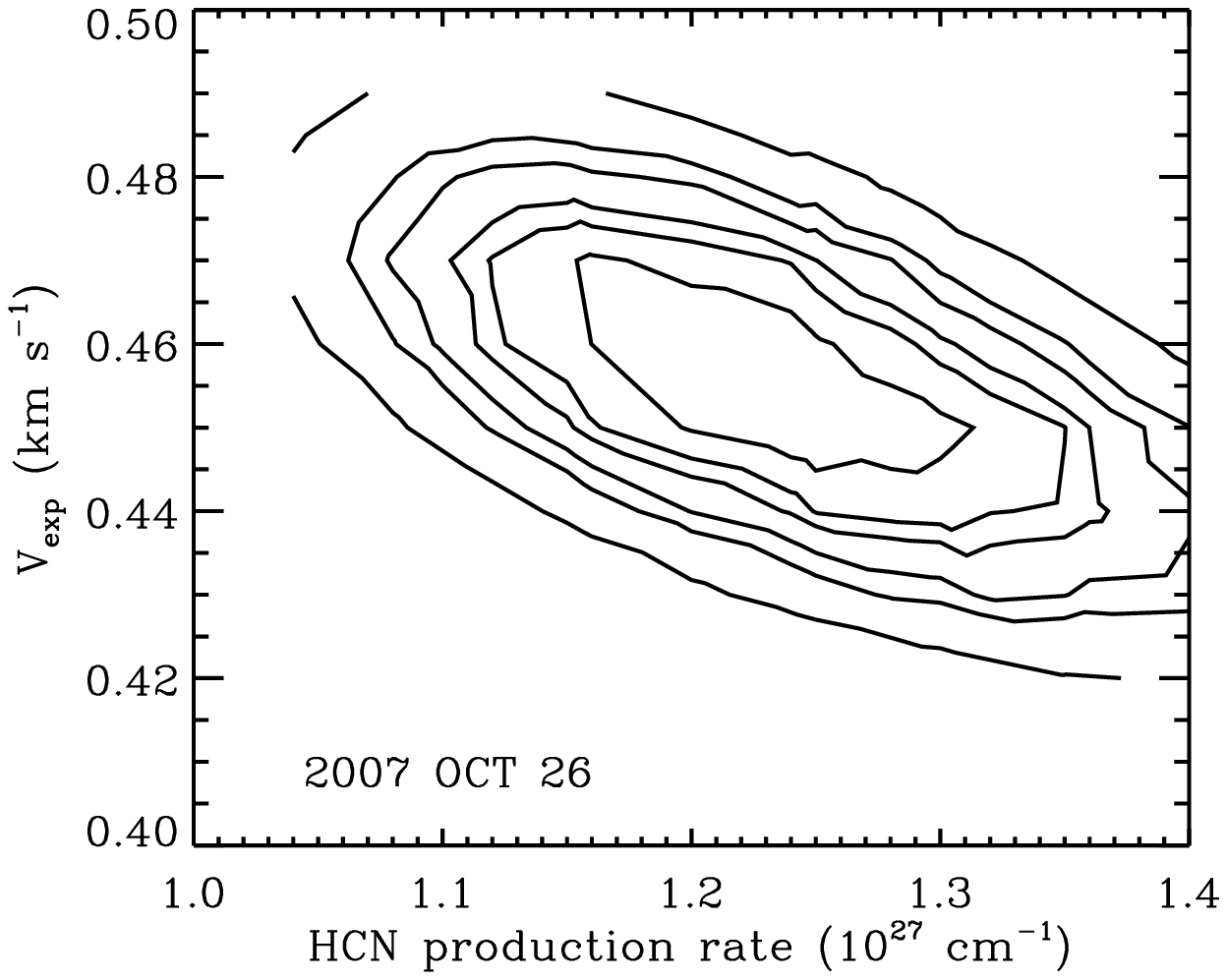} \\
\vskip 2mm
\includegraphics[width=3in]{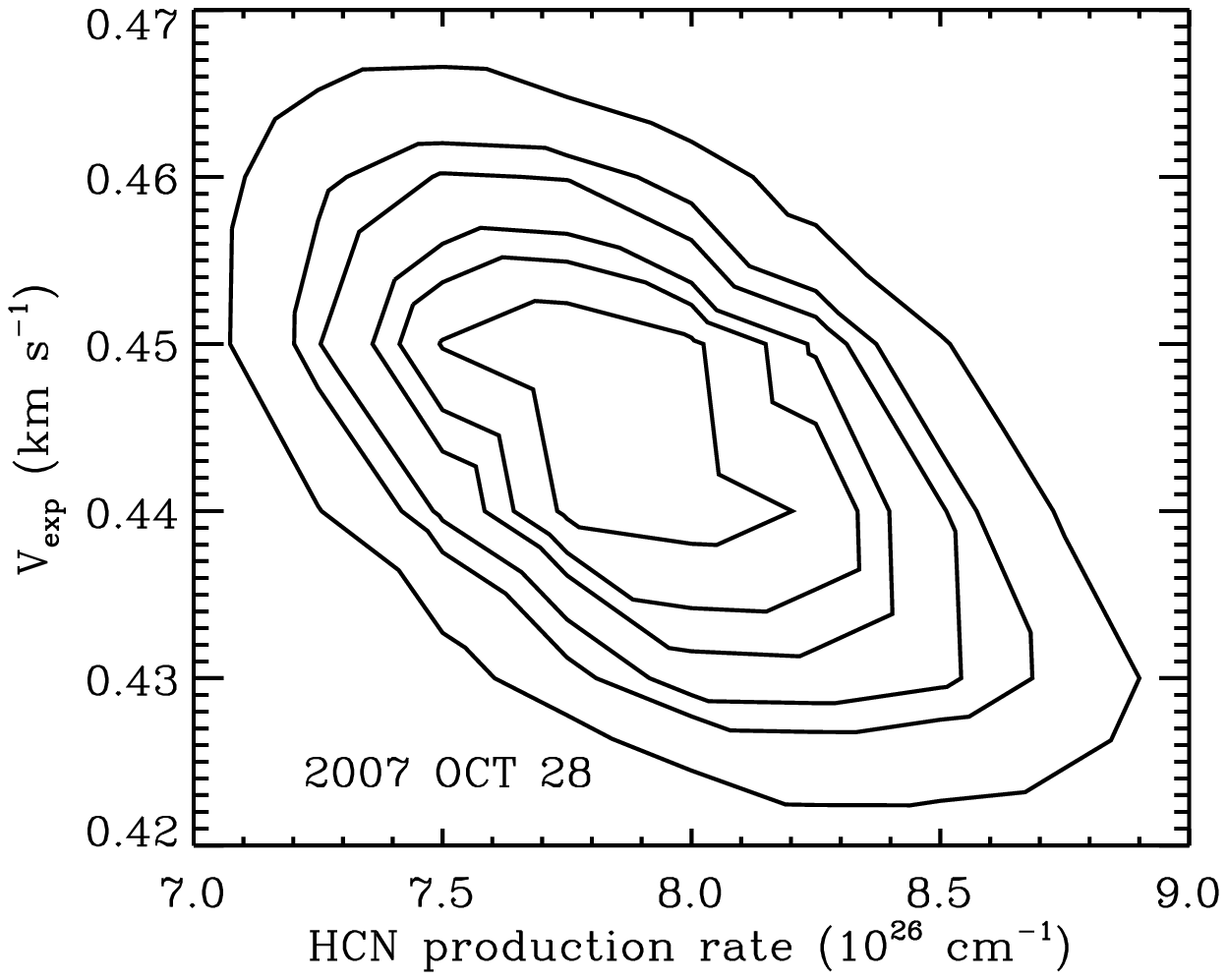} \\
\vskip 2mm
\includegraphics[width=3in]{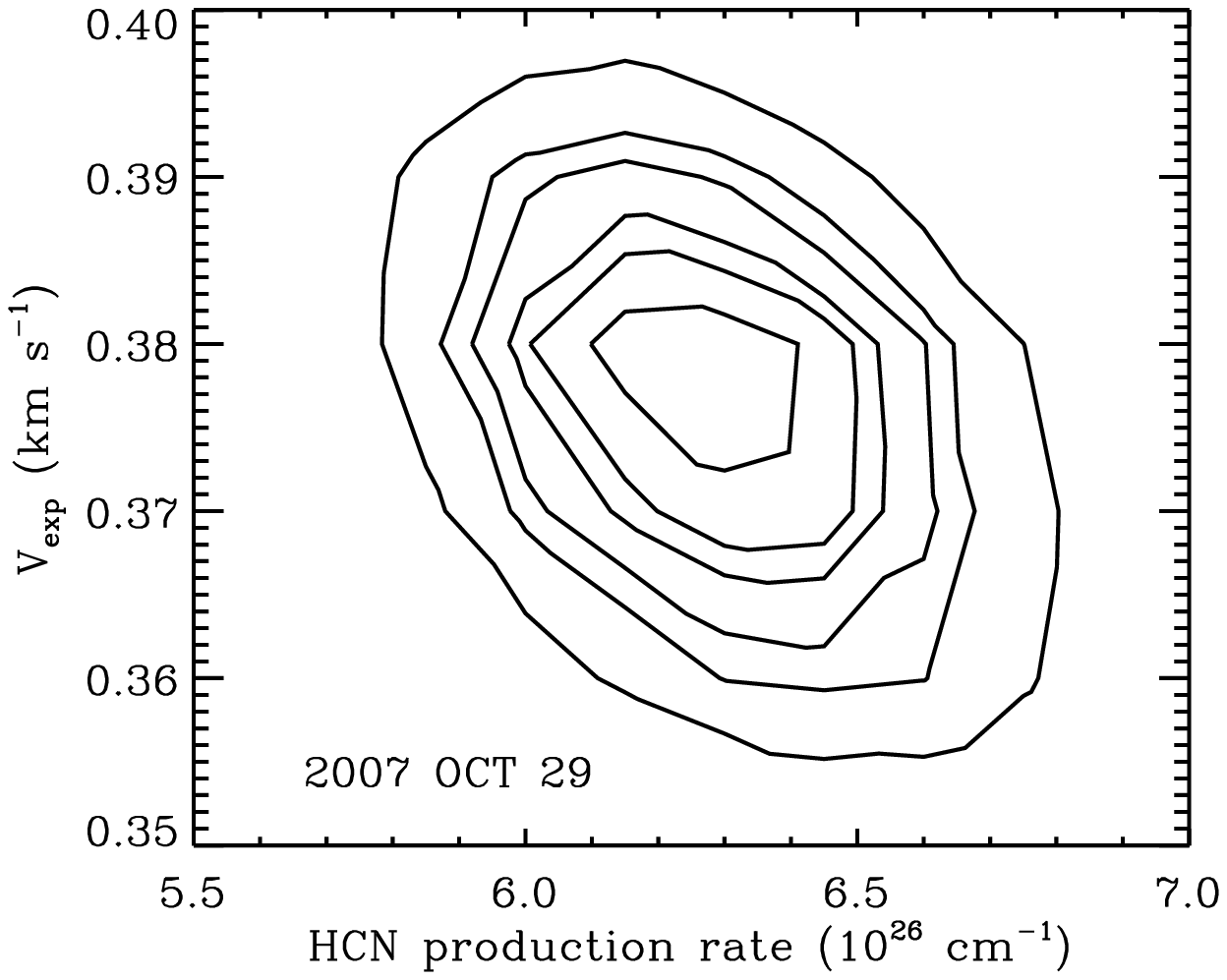}
\caption{Iso-$\chi^2$ surfaces of $Q_{\rm HCN}$ versus $v_{\rm exp}$. 
Contours correspond to the 1--6 $\sigma$ errors. \label{fig:chi2}
}
\end{figure}

\begin{figure}[t!]
\figscale{0.5}
\plotone{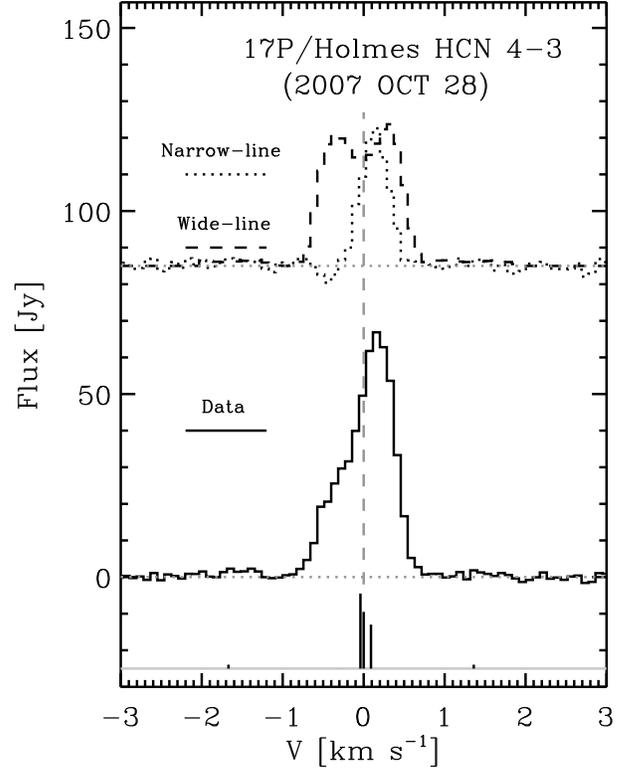}
\figcaption{ The broad-line component (dashed line) and narrow-line
  component (dotted line), compared with the full HCN data spectra
  (solid line) from 17P/Holmes on 2007 October 28 UT. The positions and
relative intensities of the hyperfine components of HCN 4--3 transition
are shown at the bottom. The
  broad-line component is retrieved from the best-fit symmetric
  outgassing model with v$_{\rm exp}$=0.44~km~s$^{-1}$ and
  Q$_{\rm HCN}$=8$\times$10$^{26}$ s$^{-1}$. The narrow-line component is
  derived from the residual of the HCN emission subtracted from the
  best-fit symmetric outgassing model. The vertical dashed line marks
  the cometocentric velocity of 0 km~s$^{-1}$. \label{fig:oct28hcn}
}

\end{figure}

The best-fit HCN production rates and expansion velocities are derived
and listed in Table~\ref{tab:fitting}.  Figures~\ref{fig:oct26},
\ref{fig:oct28}, and \ref{fig:oct29} present comparisons between the
observed channel maps and the best-fit model.  The model reproduces
the main features of symmetric pattern of the HCN
emission. The line optical depth at center is around 0.5--0.6, consistent with
the optical depth assessed above using the hyperfine components ratio. 
Figure~\ref{fig:chi2} shows the $\chi^2$ surfaces for the
Q$_{\rm HCN}$ versus V$_{\rm exp}$ values, which enables us to
quantify the uncertainties associated with the model production rates
and the expansion velocities fit for the three dates.  The best-fit
HCN production rates decrease from the 26th to the 29th
and the outflow expansion velocity drops quickly.  However there are
significant residuals between the data and model.
Figure~\ref{fig:oct28hcn} shows the spectra of the symmetric HCN line
and the residual, compared with the full spectra of HCN line taken on
October 28, demonstrating the narrow line width of the residual compared
to the symmetric model.

Using this symmetric model, we need Q$_{\rm
  HCN}$ = (1.20$\pm$0.01)$\times$10$^{27}$ and Q$_{\rm
  H^{13}CN}$ = (1.5$\pm$0.3) $\times$10$^{25}$ s$^{-1}$ to match the October 28
HCN and H$^{13}$CN 4--3 integrated intensities (see
Table~\ref{tab:line}). Therefore we derive H$^{12}$CN/H$^{13}$CN = 80$\pm$16,
which is consistent with the nominal terrestrial value of 89 and with the
measurement (H$^{12}$CN/H$^{13}$CN=114$\pm$26) by Bockel\'ee-Morvan et
al. (2008). We note that the production rates reported in Table 4 are
derived to fit only the part of the HCN lines contributed from the
symmetric outgassing. It is not possible
to fit the full HCN and H$^{13}$CN emission 
using only 
the symmetric model. However, as long as we can generate similar integrated intensities
and treat both lines in the same way, this should give reasonable
estimates of the abundance ratio. With the same method, we obtain Q$_{\rm
  HCN}$ = 1.5$\times$10$^{27}$ for 2007 October 27 by interpolating the
HCN production rates between October 26 and October 28, and the upper limit of
Q$_{\rm DCN}$ = 1.2$\times$10$^{25}$ to achieve 0.42~Jy~km~s$^{-1}$
flux density (3$\sigma$ upper limit) of DCN 3--2 emission obtained on
2007 October 27, and derive DCN/HCN $<$ 8$\times$10$^{-3}$, consistent with the bulk
D/H ratio in HCN measured from comet Hale-Bopp (2.3$\pm$0.4
$\times$10$^{-3}$, Meier et al.~1998). However, we cannot yet separate
the contribution from the freshly released material in the nucleus as
in ~\citet{Blake99} due to the limited sensitivity. 

The derived bulk HCN production rate of 1.2 $\times$ 10$^{27}$
s$^{-1}$ on October 28 and an interpolated value of 1.5 $\times$ 10$^{27}$
s$^{-1}$ on October 27 are generally consistent with the values derived
from single dish observations (2.0$\times$ 10$^{27}$, IRAM 30m,
Bockel\'ee-Morvan et al.~2008) and infrared observations (2.4$\times$
10$^{27}$, NIRSPEC, Dello Russo et al.~2008) derived on October 27,
considering the uncertainties involved in the measurements and the use of 
different aperture sizes.

\subsection{Narrow-line component}
\begin{figure}
\figscale{0.5}
\plotone{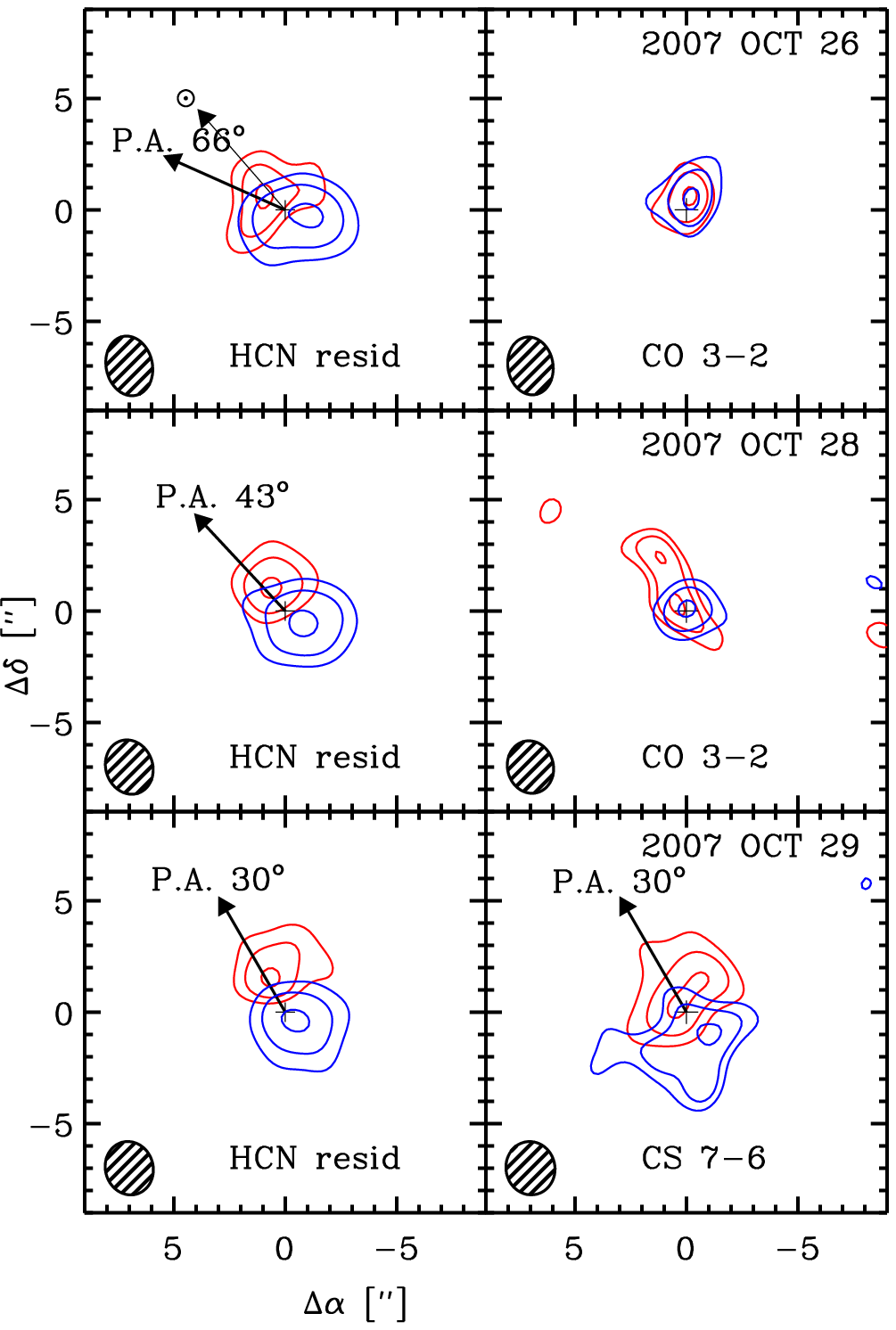}
\figcaption{Integrated intensity images red-shifted (integrated from 0.3
  to 0.6 km~s$^{-1}$) vs blue-shifted (integrated from $-$0.3 to 0
  km~s$^{-1}$) for HCN 4--3 residuals, CO 3--2 and CS 7--6 on 2007 October
  26--29. The contours are at 50\%,70\%, and 95\% of the peak
  value. The cross marks the peak of the continuum emission, which
  locates the position of the nucleus. P.A. is measured from the peak
  of the blue-shifted image to that of the red-shifted image. The Sun
  direction is shown for 2007 October 26 and the solar direction only
  changed 4.1 degrees clockwise from October 26 to 29. \label{fig:pa}
}
\end{figure}

As shown in Figure~\ref{fig:oct28hcn}, the linewidth of the
residual HCN emission is much smaller than those in the symmetric
outgassing models. The spatio-kinematic characteristics of the
residual HCN emission are very similar to those of the CO 3--2 and CS
7--6 observed simultaneously on those days. We will refer to this
component as the narrow-line component. The main characteristics that
separate it from the broad-line emission are:
\begin{enumerate}
\item The narrow-line component occupies a velocity range from $-0.2$
  to +0.4 km~s$^{-1}$, much narrower than the 
  broad-line component associated with symmetric outgassing that
  emits from $-0.8$ to +0.8 km~s$^{-1}$.
\item The line center (where the image peak coincides with the peak of
  the mm continuum emission or nucleus position) for the HCN 4--3
  residuals, CO 3--2, and CS 7--6 are all around 0.1--0.2 km~s$^{-1}$
  red-shifted, while the line
  center velocity of the broad-line component is 0 km~s$^{-1}$. 
\item The narrow-line component shows emission that is red-shifted
  toward the northeast and blue-shifted toward the
  southwest relative to the mm continuum peak. Figure~\ref{fig:pa}
  shows the red-shifted (integrated from $+$0.3 to $+$0.6 km~s$^{-1}$) and
  the blue-shifted (integrated from $-$0.3 to 0 km~s$^{-1}$) emission
  of the HCN 4--3 residuals, compared with the CO 3--2 and CS 7--6
  emission on 2007 October 26--29. Especially on the 29th, the
    position angle (P.A.) of the HCN residual is the same as that of
  CS 7--6. The velocity shift of CO 3--2 is not as
  obvious on 26th but it does show a similar extension as the HCN
  residual on 28th.
\end{enumerate}

The volatiles in the narrow-line component likely share the same
origin in the outbursting coma of 17P/Holmes. The P.A. of the red/blue
gradient of the narrow emission changed gradually from 66 to
30$^\circ$ during the period of observations. Although in the same
sense, this is a much larger shift than that of the P.A. of the
line toward the Sun, which only changed from 41.0 to 36.9$^\circ$.
We propose that the (residual) HCN, CO, and CS emission of the narrow
component may 
come from volatiles released from very small grains of pure ice
freshly released from the nucleus.  Without absorptive dust, the
ice warms up at a much slower rate.  CO might be released
faster than HCN and CS (presumably a product of CS$_2$ in the ice),
accounting for the morphological difference between CO and CS (and the
HCN residuals). We hypothesize that the narrow-line component
originates from the ice grain halo found in the near-nucleus
photometry which is believed to be created by sublimating ice grains
around the nucleus due to the outburst (Stevenson \& Jewitt
2012). Based on time-series photometry of 17P/Holmes before the
outburst, Snodgrass et al.~(2006) estimated a rotation period for
17P/Holmes of $7.2<P_{rot}<12.8$ h. If the change of the P.A. of the
narrow-line component was caused by the rotation of the nucleus, the
gradual change from 66 to 30 degrees from 26th to 29th implies a
nucleus rotational period of 11.8 h (clockwise in the plane
perpendicular to the line of the sight) or 12.2 h (anti-clockwise). We
discuss the nature of this narrow-line component further in the
discussion section.

\subsection{CO/HCN ratios}
\begin{figure}[htbp]
\figscale{0.7}
\plotone{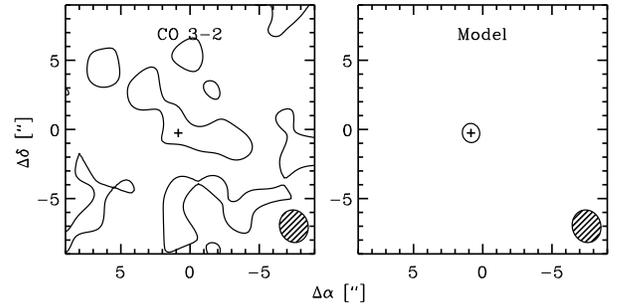}
\figcaption{ {\it left}: The integrated intensity map of CO 3--2 toward
  17P/Holmes on 2007 October 28 UT, integrated between ($-$0.8,$-$0.2) and
  (0.4,0.8) km~s$^{-1}$; {\it right}: the simulated map with
  v$_{\rm exp}$=0.44~km~s$^{-1}$ and Q$_{\rm CO}$=5.0$\times$10$^{27}$ s$^{-1}$,
  integrated over the same velocity range. The contour level is 0.08
  Jy~Beam$^{-1}$~km~s$^{-1}$. The peak of the simulated
  map matches the rms (0.09 Jy~Beam$^{-1}$~km~s$^{-1}$) of the left
  image, providing the upper limit of the contribution to the CO
  emission from the broad-line component.  \label{fig:oct28co}  
}

\end{figure}

\begin{figure}[htbp]
\figscale{0.5}
\plotone{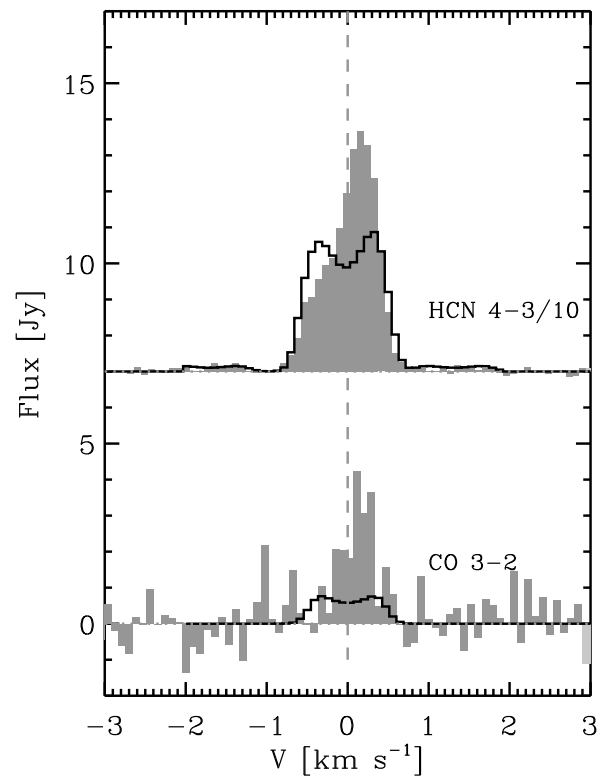}
\figcaption{Broad-line components of HCN 4--3 and CO 3--2 emission (upper
  limit) from 17P/Holmes on 2007 October 28 UT. The solid-line HCN
  spectra, retrieved from the best-fit symmetric outgassing model with
  v$_{\rm exp}$=0.44 km~s$^{-1}$ and Q$_{\rm HCN}$=8$\times$10$^{26}$~s$^{-1}$, is
  overlaid with the data spectra in gray. The solid-line CO spectra is
  the simulated spectra from the model with v$_{\rm exp}$=0.44~km~s$^{-1}$
  and Q$_{\rm CO}$=5$\times$10$^{27}$ s$^{-1}$. The simulated CO emission
  model generates the intensity image integrated between (-0.8,-0.2)
  and (0.4,0.8) km~s$^{-1}$ as shown in Figure~\ref{fig:oct28co},
  giving the upper limit of the contribution to the CO emission from
  the broad-line component. The vertical dashed line marks
  the cometocentric velocity of 0 km~s$^{-1}$. \label{fig:wide}
}
\end{figure}

\begin{figure}[t]
\figscale{0.5}
\plotone{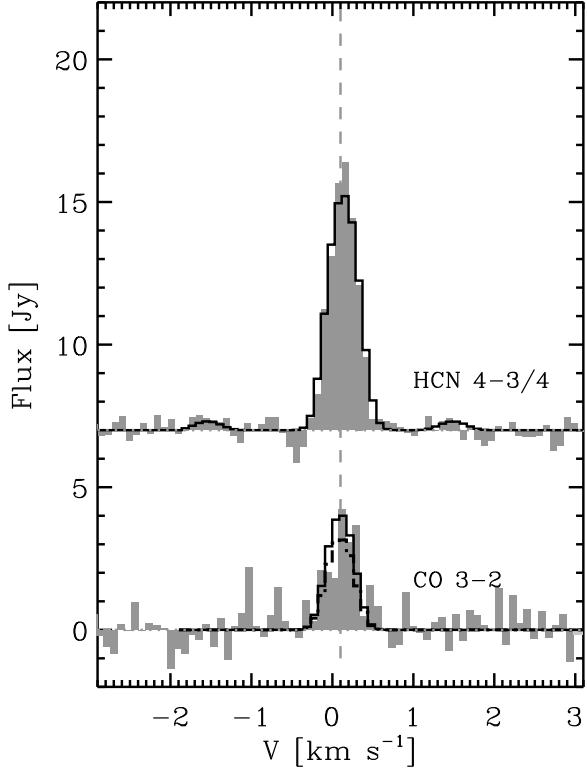}
\figcaption{Narrow-line components of HCN 4--3 and CO 3--2 emission from
  17P/Holmes on 2007 October 28 UT.  The HCN spectra (in gray) is
  retrieved from the residual of HCN emission subtracted by the
  best-fit symmetric outgassing model with v$_{\rm exp}$=0.44~km~s$^{-1}$ and
  Q$_{\rm HCN}$=8$\times$10$^{26}$ s$^{-1}$. The solid-line HCN
  spectra is
  the simulated spectra from the model with v$_{\rm exp}$=0.15~km~s$^{-1}$ and
  Q$_{\rm HCN}$=1$\times$10$^{26}$ s$^{-1}$ which generates the same flux
  as the spectra in gray. The solid-line CO spectra is simulated from
  the model with v$_{\rm exp}$=0.15~km~s$^{-1}$ and
  Q$_{\rm CO}$=4.5$\times$10$^{27}$~s$^{-1}$ which matches the CO data flux in
  the gray spectra, 1.75~Jy~km~s$^{-1}$, while the dotted-line CO
  spectra is simulated with v$_{\rm exp}$=0.15~km~s$^{-1}$ and
  Q$_{\rm CO}$=3.5$\times$10$^{27}$~s$^{-1}$ which generates the
  integrated flux 1.35~Jy~km~s$^{-1}$. The vertical dashed line marks
  the cometocentric velocity of $+$0.1 km~s$^{-1}$. \label{fig:resid}
}
\end{figure}

Here we investigate the CO/HCN ratios within the two components based
on the data obtained on 2007 October 28, which had the
  finest spectral resolution for both HCN 4--3 and CO 3--2.  For HCN,
the parameters of the broad-line component come from
the best-fit symmetric outgassing model as described above, by fitting
the data avoiding the velocity range from $-$0.2 to +0.4
km~s$^{-1}$. The best-fit production rate is 8.0$\times$
10$^{26}$~s$^{-1}$ and the expansion velocity 0.44 km~s$^{-1}$. The
narrow-line component is retrieved from the residual HCN emission
after subtraction of the best-fit symmetric outgassing
model. Figure~\ref{fig:oct28} and \ref{fig:oct28hcn} show the
simulated HCN images and spectra of the two components compared with
the data.  As shown in the figures the narrow-line component mainly
emits over the velocity range of $-$0.2 to +0.4 km/s, which is
consistent with the CO emission obtained simultaneously. This
indicates that most of the CO emission, if not all, must originate
from the narrow-line component.

In order to estimate how much the broad-line
component contributes to the CO emission, we make use of the emission
avoiding the velocity range from $-$0.2 to +0.4 km~s$^{-1}$, similar
to the HCN analysis. However, since the signal is too weak for the CO
emission, we integrate the emission with velocity between $-$0.8 to
+0.8 km~s$^{-1}$, again avoiding the velocity range from $-$0.2 to
+0.4 km~s$^{-1}$, which is clearly dominated by the narrow-line
component. Figure~\ref{fig:oct28co} shows the integrated intensity
image, which does not show any significant emission left with the
noise level at 0.09 Jy~Beam$^{-1}$~km~s$^{-1}$. The upper limit of the
broad-line component contribution can be obtained by
simulating CO emission with the same broad-line
expansion velocity as determined in HCN emission, i.e., 0.44
km~s$^{-1}$, and matching the 1$\sigma$ integrated flux level within
the same velocity range.  This is achieved with Q$_{\rm CO}$
= 5.0$\times$10$^{27}$~s$^{-1}$. Figure~\ref{fig:oct28co} compares the
integrated intensity of the simulated image with that of CO data and
Figure~\ref{fig:wide} demonstrates the broad-line components of HCN
4--3 and CO 3--2, compared with the data. The CO/HCN ratio in
the broad-line component is therefore $<$ 7. Note that the upper limit
of the CO 3--2 broad-line component contributes 0.3 Jy~km~s$^{-1}$ in
the velocity range from $-$0.2 to 0.4 km~s$^{-1}$, which is about 1/6
of the total integrated intensity of CO 3--2 emission.

In order to obtain the CO/HCN ratio in the narrow-line component, we
first simulate HCN emission based on the HCN residual, using the same
symmetric outgassing model described above, only with much smaller
expansion velocity. However we only fit on the integrated flux based
on the spectra, as we cannot reproduce the spatial shift across the
nucleus position. Although not ideal, this approach should give a
reasonable estimate of the abundance ratio, as long as CO and HCN are
treated in the same way. Fitting the residual HCN spectra yields a
production rate of Q$_{\rm HCN}$ = 1.0$\times$10$^{26}$~s$^{-1}$ and
expansion velocity 0.15~km~s$^{-1}$. Assuming there is no contribution
of the broad-line component to the CO line, we find
Q$_{\rm CO}$ = 4.5$\times$10$^{27}$~s$^{-1}$ to match the integrated
intensity of 1.75 Jy~km~s$^{-1}$ for CO 3-2 with an expansion velocity
of 0.15~km~s$^{-1}$. This results in a CO/HCN ratio of 45. If we
consider a maximum contribution of the wide component to the CO line
equal to the upper limit derived above, i.e., 0.3 Jy~km~s$^{-1}$,
we find Q$_{\rm CO}$ = 3.5$\times$10$^{27}$~s$^{-1}$ to reproduce the
integrated intensity of 1.45 Jy~km~s$^{-1}$, and a CO/HCN ratio =
35. Figure~\ref{fig:resid} shows the narrow-line components of HCN
4--3 and CO 3--2, compared with the CO data and residual HCN spectra.
Since the unknown contribution from the broad-line
component dominates the uncertainty of the calculation, we conclude
that the ratio of CO/HCN in the narrow-line component can be
constrained to 40$\pm$5. This is significantly higher than the
CO/HCN$<7$ found for the broad-line component. 
Using the same analysis, we determine the ratio CO/HCN $<$ 9 for the
broad-line component and 45$\pm$10 for the narrow-line component on
October 26. This is consistent with the result on October 28, despite larger
uncertainties due to the fact that the spectral resolution of HCN and CO lines
carried out on October 26 is a factor of two worse than on October 28, which 
creates large uncertainties in how much the broad-line component contributes
to the CO emission on October 26.

HCN is the molecule which exhibits the lowest abundance
variation from comet to comet (Biver et al.~2002; Mumma \& Charnley
2011). Therefore, production rates relative to that of
HCN can be used for a comparative study of molecular
abundances. From a study of a sample of 24 comets, Biver et al.~(2002) 
found that CO/HCN varies from $<$24 to 180.
In particular, 4 long period comets show CO/HCN$>$85 while 4 other comets,
including one short period comet, show  CO/HCN$<$40.
However, there is still no obvious correlation
between the dynamical class and the chemical composition of comets
(Biver et al.~2002; Crovisier, 2007). We find in the near-nucleus
outgassing of comet 17P/Holmes, the CO/HCN ratio is much higher in 
the narrow component than the broad one. 
The difference of CO/HCN ratios in the two components
demonstrates the heterogeneity of the nucleus of this Jupiter Family
Comet.

The decrease in the HCN production rates indicates an $e-$folding time $\sim$4 days in the 
October 26 to 29 period (Table 4).  Schleicher (2009) measured the
decline of the production rates of optical species. From their Table 3
we find an  
$e-$folding time $\sim$14 days in the 2007 November to 2007 December
period, with an indication that the fading timescale increases as time
passes up to their last observation in 2008 March. 
Interestingly, their CN production rate, $\log_
{10} (Q_{\rm CN})$ = 27.4 on November 1, is higher than the HCN production
rate $\log_{10} (Q_{\rm HCN})$ = 27.1 we derived four days earlier (October 28) when matching
the whole flux of HCN 4--3 emission. This inequality
could be an artifact of the use of the Haser model (which applies to a
coma expanding in steady state and which may not accurately describe
the outburst coma of 17P) or it could be an indication that CN has a
parent in addition to HCN. Furthermore, the photometer entrance apertures used in the optical
measurements, ranging from 24 to 204 arcsec, are different from the
primary beams (31 arcsec) used in our observations. This
could also contribute to the difference in production
rates derived for the CN and HCN molecules.

\section{Discussion}

What are the two components in the near-nucleus coma of the 
outbursting 17P/Holmes?
As shown from the above model fitting, the HCN emission from the broad-line
component can be fit with symmetric outgassing with an expanding
velocity about 0.5 km~s$^{-1}$. Drahus et al.~(2007) and Snodgrass et
al. (2007) found that an almost-spherical dust coma expanded around
the nucleus of 17P/Holmes at 0.55 km~s$^{-1}$.  Both the expansion 
velocity and symmetric outgassing pattern of the broad-line component 
appear to be related to this spherical dust coma releasing dust from the 
surface of the nucleus.

However, within the dust coma, bright streaks emerged from the
nucleus, likely produced by fragments that moved away
from the nucleus with velocities of up to 0.1 km~s$^{-1}$
(Trigo-Rodriguez et al.~2007; Stevenson et al.~2010). Gaillard et al.~(2007) also postulated
that the streaks originated from disintegrating cometary fragments.
The images from Hubble telescope's Wide
Field Planetary Camera 2 (WFPC2) also showed a ``bow tie''
appearance\footnote{http://www.spacetelescope.org/news/heic0718/},  
about twice as much dust
lies along the east-west direction as along the north-south direction,
similar to the morphology we have detected in the narrow-line
component. We find the production rate from the broad-line component of HCN 4--3
decreased but the whole integrated intensity of the line actually
increased from October 28th to 29th, which implies that the
narrow-line component should have increased. Coupled with the
increase of the mm continuum flux level between the two days, this 
provides strong evidence for disintegrating cometary fragments 
continuously supplying the dust and gas near the nucleus.
So we believe the narrow-line component
detected in the SMA spectral line imaging shown here should
associate with volatiles sublimating from the
fragments near the nucleus which  was the most recently ejected and
most pristine in nature. 
The main problem with this interpretation is how long these streaks/fragments
persisted. Stevenson \& Jewitt (2012) presented the near-nucleus
photometry over a period of three months following the outburst of
comet 17P/Holmes and they found a halo consisting of freshly released
icy grains persisted around the nucleus and extended to radii of
several thousand kilometers, encompassing the emission area we
detected. In our interpretation, the narrow-line component
likely originated from this halo, with volatiles continuously supplied
from the freshly released icy grains.  Yang et al.~(2009) suggested
that only clean ice grains, with $<$10\% impurities, could survive long enough
to travel 3500 km from the nucleus before sublimating, which
  could explain the offsets from the nucleus seen in the narrow-line 
component as in emission of CO, CS and HCN residuals.
The line width of the narrow-line component does not change much
with time over the four days of observations, compared to the rapid narrowing of the
expansion velocity of the symmetric outgassing component. This is
consistent with the long lifetime of the halo feature found by Stevenson
\& Jewitt (2012).

\section{Summary}

The Jupiter family comet 17P/Holmes underwent a dramatic outburst in
2007,  providing us with a rare  opportunity to
access fresh material released by the powerful outburst from beneath
the thermally-processed
surface. We have observed 
17P/Holmes with
the SMA from October 26 through October 29 2007 at a spatial
resolution of $\sim$3000 km at the comet distance and found peculiar
outgassing in the near-nucleus coma. Our key results include the
following.  

\begin{enumerate}

\item We find two components in the molecular emission from the
  near-nucleus coma: one has relatively broad linewidth ($\sim$1 km
  s$^{-1}$ FWHM), showing a symmetric outgassing pattern with respect
  to the nucleus position. The other has a narrow linewidth ($<$0.5 km
  s$^{-1}$ FWHM). HCN 4--3 emission has both components, while the
  emission of CS 7--6 and CO 3--2 is dominated by the narrow-line
  component.   

\item The morphologies of the two components are different: The
  peaks of the broad-line component in the HCN 4--3 emission are 
  on the nucleus position and do not change with velocity;  the
  line center of the narrow-line component is red-shifted by 0.1--0.2
  km s$^{-1}$ (cometocentric frame). The peaks of the emission shift
  with velocity across the nucleus position and the position angle
  of   the red/blue gradient changed from $66\degr$ to $30\degr$
  during the four days of observations. 

\item We determine distinctly different CO/HCN ratios for each of the
  components: CO/HCN $<$7 in the broad-line component and
  CO/HCN=40$\pm$5 in the narrow-line one. This different line ratio
  reflects the heterogeneity of the nucleus of this Jupiter family
  comet.  

\item We determine the $^{12}\rm C$/$^{13}\rm C$ ratio=80$\pm$16 based on the
  simultaneous observations of HCN and H$^{13}$CN 4--3 emission, which
  is consistent with the terrestrial value.

\end{enumerate}

These observations demonstrate the unique power of radio interferometers with 
high spatial and spectral resolution to map small scale
spatio-kinematic differences in emission
of primary volatiles in the near-nucleus cometary comae to probe the 
heterogeneity in the nucleus. Simultaneous observations of multiple
primary volatiles are key to understanding and interpreting the 
outgassing properties of the nucleus. The newly constructed 
Atacama Large Millimeter/Submillimeter Array, or ALMA, will
allow to assess the composition of a large sample of comets to examine
its relationship with cometary formation region.

{\it Facilities:} \facility{SMA}

\acknowledgements
\noindent  The SMA is a joint project between the Smithsonian
Astrophysical Observatory and the Academia Sinica Institute of
Astronomy and Astrophysics and is funded by the Smithsonian
Institution and the Academia Sinica. We thank SMA Director Raymond
Blundell for quick allocation of the DDT time for the above
observations and the SMA staff, especially Ken (Taco) Young  and Nimesh
Patel, for observational supports.   DJ acknowledges support from NASA's Origins program.


\end{document}